\documentclass[AMA,STIX2COL]{MRM}
\usepackage{setspace}
\usepackage{amssymb}
\usepackage{amsmath}
\usepackage{multirow}
\usepackage{comment}
\usepackage{lipsum}  
\usepackage{graphicx}
\usepackage{pgffor} 
\usepackage{etoolbox} 
\usepackage{xstring} 
\usepackage{booktabs}
\usepackage{amsbsy} 
\usepackage{siunitx}
\usepackage{xparse}
\usepackage[outline]{contour}
\usepackage{calc}

\usepackage{color,pxfonts,fix-cm}
\usepackage{latexsym}
\usepackage[mathletters]{ucs}
\usepackage[T1]{fontenc}
\usepackage[utf8x]{inputenc}
\usepackage{pict2e}
\usepackage{wasysym}
\usepackage{anyfontsize}
\usepackage{tikz}
\usepackage{xcolor}

\usepackage{calc}
\usepackage{xspace}

\usepackage{chngcntr}
\usepackage{caption}
\usepackage{newfloat}

\renewcommand{\eqref}[1]{Equation~(\ref{eq:#1})}

\newcommand{\tabref}[1]{Table~\ref{tab:#1}\!\!\!}
\newcommand{\figref}[1]{Figure~\ref{fig:#1}\!\!\!} 
\newcommand{\figsupref}[1]{Figure~S\ref{fig:#1}} 

\newcolumntype{L}[1]{>{\raggedright\let\newline\\\arraybackslash\hspace{0pt}}m{#1}}
\newcolumntype{C}[1]{>{\centering\let\newline\\\arraybackslash\hspace{0pt}}m{#1}}
\newcolumntype{R}[1]{>{\raggedleft\let\newline\\\arraybackslash\hspace{0pt}}m{#1}}

\DeclareMathOperator*{\argmin}{arg\,min}

\renewcommand{\Hat}{\widehat}

\renewcommand{\vec}[1]{\ensuremath{\boldsymbol{#1}}}

\newcommand{\Hvec}[1]{\ensuremath{\Hat{\boldsymbol{#1}}}}
\newcommand{\tvec}[1]{\ensuremath{\tilde{\boldsymbol{#1}}}}

\newcommand{\Real}{{\mathbb{R}}}

\newcommand{\moco}{NRIR\xspace}
\newcommand{\vxm}{VxM\xspace}
\newcommand{\vxmcc}{VxM-CC\xspace}
\newcommand{\vxmed}{VxM-ED\xspace}
\newcommand{\aimcc}{AiM-CC\xspace}
\newcommand{\aimed}{AiM-ED\xspace}

\newcommand{\imagetextsize}{\fontsize{9pt}{12.5pt}\selectfont}

\DeclareFloatingEnvironment[
    fileext=los,
    listname={List of Supplementary Figures},
    name=Figure,
    placement=htbp,
    within=none
]{suppfigure}
\DeclareCaptionLabelFormat{suppfig}{Figure S#2~}
\articletype{RESEARCH ARTICLE}
\received{xx Aug 2024}
\revised{xx Dec 2024}
\accepted{xx Feb 2025}
\begin{document}

\title{Groupwise Image Registration with Edge-Based Loss for Low-SNR Cardiac MRI}

\author[1]{Xuan Lei}{\orcid{0009-0009-0878-240X}}
\author[1]{Philip Schniter}{\orcid{0000-0003-0939-7545}}
\author[2]{Chong Chen}{\orcid{0000-0002-4965-5564}}
\author[2,3]{Rizwan Ahmad}{\orcid{0000-0002-5917-3788}}

\authormark{Lei \textsc{et al.}}

\address[1]{\orgdiv{Electrical \& Computer Engineering}, \orgname{The Ohio State University}, \orgaddress{\state{Ohio}, \country{USA}}}

\address[2]{\orgdiv{Biomedical Engineering}, \orgname{The Ohio State University}, \orgaddress{\state{Ohio}, \country{USA}}}

\address[3]{\orgdiv{Davis Heart and Lung Research Institute}, \orgname{The Ohio State University Wexner Medical Center}, \orgaddress{\state{Ohio}, \country{USA}}}

\corres{Rizwan Ahmad, \email{ahmad.46@osu.edu}}
\presentaddress{Biomedical Research Tower, 460 W 12th Ave, Room 318, Columbus OH 43210, USA.}

\finfo{\fundingAgency{National Heart, Lung, and Blood Institute} Grant Numbers: \fundingNumber{R01EB029957} and \fundingNumber{R01HL151697}}

\abstract[Abstract]{
\section{Purpose} To perform image registration and averaging of multiple free-breathing single-shot cardiac images, where the individual images may have a low signal-to-noise ratio (SNR).

\section{Methods} To address low SNR encountered in single-shot imaging, especially at low field strengths, we propose a fast deep learning (DL)-based image registration method, called Averaging Morph with Edge Detection (\aimed). \aimed jointly registers multiple noisy source images to a noisy target image and utilizes a noise-robust pre-trained edge detector to define the training loss. We validate \aimed using synthetic late gadolinium enhanced (LGE) images from the MR extended cardiac-torso (MRXCAT) phantom and free-breathing single-shot LGE images from healthy subjects (24 slices) and patients (5 slices) under various levels of added noise. Additionally, we demonstrate the clinical feasibility of \aimed by applying it to data from patients (6 slices) scanned on a 0.55T scanner.

\section{Results} Compared to a traditional energy-minimization-based image registration method and DL-based VoxelMorph, images registered using \aimed exhibit higher values of recovery SNR and three perceptual image quality metrics. An ablation study shows the benefit of both jointly processing multiple source images and using an edge map in \aimed. 

\section{Conclusion} For single-shot LGE imaging, \aimed outperforms existing image registration methods in terms of image quality. With fast inference, minimal training data requirements, and robust performance at various noise levels, \aimed has the potential to benefit single-shot CMR applications. }

\wordcount{3898}

\jnlcitation{\cname{%
\author{X. Lei}, 
\author{P. Schniter},
\author{C. Chen},
\author{R. Ahmad}} (\cyear{2025}), 
\ctitle{Groupwise image registration with edge-based loss for low-SNR cardiac MRI}, \cjournal{Magn. Reson. Med.}}

\maketitle
\clearpage
\section{Introduction}\label{sec:intro}
Cardiovascular magnetic resonance imaging (CMR) is an MRI-based technique that provides a comprehensive assessment of the cardiovascular system. 
A single CMR exam can provide functional, structural, and morphological assessment of the heart. 
CMR is considered a gold standard for assessing biventricular cardiac function and myocardial viability. \cite{pennell2002ventricular}  
The former is typically assessed using cine, while the latter is assessed using late gadolinium enhancement (LGE). \cite{elliott2005late}

Long scan times and the requirement for several breath-holds remain significant challenges for patients, particularly those who are ill or have difficulty remaining still for extended periods. 
As a consequence, free-breathing real-time imaging for cine and flow \cite{nayak2005future} and free-breathing single-shot imaging for LGE \cite{piehler2013free} and parametric mapping \cite{kellman2015free} are being increasingly utilized in clinical settings, as they significantly reduce scan times and eliminate the need for breath-holds. Improving the quality of accelerated real-time or single-shot imaging has been an active area of research and development. \cite{vermersch2020compressed, zucker2021free}

In single-shot LGE imaging, an image is acquired in the diastolic phase every one to two heartbeats. \cite{kellman2002phase, kellman2016free} 
To improve signal-to-noise ratio (SNR), a common practice is to collect and average multiple single-shot images. Due to the respiration-induced cardiac motion, the single-shot images are not aligned and thus can only be averaged after image registration. 
The quality of the final registered image depends on the SNR of individual images, the number of images, and the accuracy of the registration process. 
With the increasing commercial availability of low-field scanners, \cite{campbell2024cardiac, qin2022sustainable} which suffer from low SNR, the importance of effective image registration becomes even more critical to ensure that the averaged image achieves the necessary diagnostic quality.

In thoracic imaging, rigid and affine motion models do not adequately characterize the respiration-induced motion of the heart and the surrounding organs. Therefore, cardiopulmonary motion is often modeled as deformable motion. \cite{amelon2011three, mcleish2002study}
Traditional energy-minimization-based registration methods, such as ANTS and ELASTIX, \cite{avants2009advanced, klein2009elastix} are frequently employed to handle complex cardiopulmonary motion. However, long computation times pose a major limitation for the clinical utility of these methods. More recently, deep learning (DL)-based approaches, such as VoxelMorph (VxM), \cite{balakrishnan2019voxelmorph} have been proposed for image registration. 
\vxm can be used as an unsupervised method, where a registration network is trained on a given pair of target-source images. However, this approach makes \vxm computationally slow and prone to overfitting, especially when the SNR of the target image is low. 
Fortunately, a key feature of \vxm is that the registration network can be trained using a small number of target-source image pairs and then used for image registration without further training.

In this work, we present an extension of \vxm, termed Averaging Morph with Edge Detection (\aimed), which allows groupwise registration\cite{bhatia2004consistent} to jointly register multiple source images to one target image. Two key features distinguish this work from prior art. First, \aimed is lightweight, employing an array of weight-tied \vxm networks. Consequently, \aimed not only can be trained on smaller datasets compared to other groupwise registration methods\cite{martin2020groupwise, van2020deep} but also offers fast inference. Second, by defining the loss function on DL-based edge maps rather than the original images, \aimed provides robustness against SNR and contrast changes. 
Using data from MR extended cardiac-torso (MRXCAT) phantom\cite{segars20104d, wissmann2014mrxcat} and LGE imaging at {3T, 1.5T, and 0.55T scanners}, we validate \aimed's performance under low SNR conditions, demonstrating its effectiveness in scenarios often encountered in CMR but not previously addressed in prior works.

\section{Methods}\label{sec:mat}
In this section, first, we briefly summarize \vxm for $d$-dimensional images. Second, we describe the proposed \aimed method using the notation set up for \vxm. Finally, we describe four studies using data (i) simulated from a realistic digital phantom, (ii) collected on a 3T scanner from 12 healthy volunteers, (iii) collected on a {1.5T} scanner from 5 patients, and (iv) collected on a 0.55T scanner from 5 patients. Studies (ii), (iii), and (iv) were performed on consented healthy subjects and patients, approved by the institutional review board (2005H0124, 2021H0414).

\subsection{VoxelMorph}\label{sec:vxm}
Let $\vec{t}\in \Real^{N}$ and $\vec{s}\in \Real^{N}$ be voxel values of $d$-dimensional target and source images, respectively, defined at known voxel coordinates $\vec{p}\in \Real^{dN}$. 
The goal of \vxm is to learn a deformation field $\vec{\phi}(\vec{p}; \vec{t},\vec{s})\in \Real^{dN }$ such that $\vec{s}\circ \vec{\phi}(\vec{p}; \vec{t},\vec{s})$ is aligned with $\vec{t}$ under some measure of similarity. Here, $\vec{s}\circ \vec{\phi}(\vec{p}; \vec{t},\vec{s})$ represents warping of $\vec{s}$ from the original $\vec{p}$ coordinates to the $\vec{\phi}(\vec{p}; \vec{t},\vec{s})$ coordinates.

In \vxm, image registration for a given target-source image pair $(\vec{t}, \vec{s})$ is accomplished by training a convolutional neural network (CNN), parameterized by $\vec{\theta}$, to solve the following optimization problem.
\begin{align}
  \scalebox{0.8}{%
  $\begin{aligned}
    \Hvec{\theta} = \argmin_{\vec{\theta}} \mathcal{L}_{\sf s}\left(\vec{t},~\vec{s} \circ \vec{\phi}_{\vec{\theta}}(\vec{p}; \vec{t},\vec{s})\right)
 + \lambda \mathcal{L}_{\sf r}\left(\vec{u}_{\vec{\theta}}(\vec{p}; \vec{t}, \vec{s})\right),
  \end{aligned}$%
  }
  \label{eq:vxm_1}
\end{align}
where the displacement vector field $\vec{u}_{\vec{\theta}}(\vec{p}; \vec{t},\vec{s})$ is the output of the CNN used in \vxm, with the subscript $\vec{\theta}$ indicating its dependence on the network parameters, and $\vec{\vec{\phi}_{\vec{\theta}}}(\vec{p}; \vec{t},\vec{s}) = \vec{p} + \vec{u}_{\vec{\theta}}(\vec{p}; \vec{t},\vec{s})$ is the deformation field.

The first term in \eqref{vxm_1} enforces similarity between $\vec{t}$ and $\vec{s} \circ \vec{\phi}_{\vec{\theta}}(\vec{p}; \vec{t},\vec{s})$, with the warping operation implemented using a spatial transformer network. \cite{jaderberg2015spatial} The second term in \eqref{vxm_1} enforces spatial smoothness to generate physically realistic displacement vector fields. Common choices for $\mathcal{L}_{\sf s}(\cdot, \cdot)$ include voxelwise mean squared difference, mean absolute difference, and normalized cross-correlation (CC), \cite{lewis1995fast} while a common choice for $\mathcal{L}_{\sf r}(\cdot)$ includes the $\ell_2$-norm of spatial gradients. \cite{fischer2002fast, balakrishnan2019voxelmorph} 

Although \vxm is capable of training a network for each target-source pair separately, its real advantage resides in training the network using multiple target-source pairs and then using the trained CNN to register images using one forward pass through the network. Given $M$ target-source pairs $\{(\vec{t}^{(i)}, \vec{s}^{(i)})\}_{i=1}^M$, a single CNN network can be trained by
\begin{align}
  \scalebox{0.8}{%
  $\begin{aligned}
    \Hvec{\theta}  = \argmin_{\vec{\theta}}\frac{1}{M} \sum_{i=1}^M &\mathcal{L}_{\sf s}\left(\vec{t}^{(i)},~\vec{s}^{(i)} \circ \vec{\phi}_{\vec{\theta}}^{(i)}(\vec{p}; \vec{t}^{(i)},\vec{s}^{(i)})\right) \\
    &\quad + \lambda \mathcal{L}_{\sf r}\left(\vec{u}_{\vec{\theta}}^{(i)}(\vec{p}; \vec{t}^{(i)}, \vec{s}^{(i)})\right).
  \end{aligned}$%
  }
  \label{eq:vxm_2}
\end{align}

After training, registration for any pair of test images $(\vec{t}^{({\sf test})}, \vec{s}^{({\sf test})})$ can be performed by first passing $\vec{t}^{({\sf test})}$ and $\vec{s}^{({\sf test})}$ through CNN to generate $\vec{\phi}_{\Hvec{\theta}}^{(\sf test)}(\vec{p}; \vec{t}^{(\sf test)},\vec{s}^{(\sf test)})$ and then performing warping $\vec{s}^{({\sf test})} \circ \vec{\phi}_{\Hvec{\theta}}^{({\sf test})}(\vec{p}; \vec{t}^{(\sf test)},\vec{s}^{(\sf test)})$ using the same spatial transformer network that was used during training.

\subsection{Proposed groupwise registration}\label{sec:aimed}
We first extend \vxm to jointly register $K$ noisy source images $\{\vec{s}_j\}_{j=1}^K \triangleq \vec{s}_{1:K}$ to one noisy target $\tvec{t}$. Given $M$ sets of clean reference $\vec{t}$, noisy target $\tvec{t}$, and noisy source images, i.e., $\{(\vec{t}^{(i)}, \tvec{t}^{(i)}, \vec{s}_{1:K}^{(i)})\}_{i=1}^M$, we train a single network by  
\begin{align}
  \scalebox{0.8}{%
  $\begin{aligned}
    \Hvec{\theta}  = \argmin_{\vec{\theta}}\frac{1}{M} \sum_{i=1}^M & \mathcal{L}_{\sf s}\left(\vec{t}^{(i)},~\frac{1}{K}\sum_{j=1}^K\vec{s}_j^{(i)} \circ \vec{\phi}_{\vec{\theta}, j}^{(i)}(\vec{p}; \tvec{t}^{(i)}, \vec{s}_{1:K}^{(i)})\right) \\
    &\quad + \frac{\lambda}{K}\sum_{j=1}^K \mathcal{L}_{\sf r}\left(\vec{u}_{\vec{\theta},j}^{(i)}(\vec{p}; \tvec{t}^{(i)}, \vec{s}_{1:K}^{(i)})\right),
  \end{aligned}$%
  }
  \label{eq:our_1}
\end{align}
where $\vec{\phi}_{\vec{\theta},j}^{(i)}(\vec{p}; \tvec{t}^{(i)}, \vec{s}_{1:K}^{(i)})$ and $\vec{u}_{\vec{\theta},j}^{(i)}(\vec{p}; \tvec{t}^{(i)}, \vec{s}_{1:K}^{(i)})$ represent the deformation field and displacement vector field, respectively, for the $j^{\sf th}$ source image from the $i^{\sf th}$ sample in the training set. 

Then, we modify the first term in \eqref{our_1} by computing $\mathcal{L}_{\sf s}(\cdot, \cdot)$ in a feature space rather than in the original voxel space. To this end, we utilize a pre-trained DL-based edge detector $\mathcal{E}(\cdot)$, which takes an image as input and generates an edge map of the same dimensionality as output, leading to the following optimization problem:

\begin{align}
  \scalebox{0.8}{%
  $\begin{aligned}
    \Hvec{\theta}  = \argmin_{\vec{\theta}}\frac{1}{M} \sum_{i=1}^M & \mathcal{L}_{\sf s}\left(\mathcal{E}\big(\vec{t}^{(i)}\big),~\mathcal{E}\big(\frac{1}{K}\sum_{j=1}^K\vec{s}_j^{(i)}\circ \vec{\phi}_{\vec{\theta}, j}^{(i)}(\vec{p}; \tvec{t}^{(i)}, \vec{s}_{1:K}^{(i)})\big)\right) \\
    &\quad + \frac{\lambda}{K}\sum_{j=1}^K \mathcal{L}_{\sf r}\left(\vec{u}_{\vec{\theta},j}^{(i)}(\vec{p}; \tvec{t}^{(i)}, \vec{s}_{1:K}^{(i)})\right).
  \end{aligned}$%
  }
  \label{eq:our_2}
\end{align}

After training, registration for a given set of noisy test images $(\tvec{t}^{({\sf test})}, \vec{s}_{1:K}^{({\sf test})})$ can be performed by passing $\tvec{t}^{({\sf test})}$ and $\vec{s}_{1:K}^{({\sf test})}$ to the trained CNN, the spatial transformer, and the mean (averaging) layer to generate a single registered image $\tfrac{1}{K}\sum_{j=1}^K\vec{s}_j^{(\sf test)}\circ \vec{\phi}_{\Hvec{\theta}, j}^{(\sf test)}(\vec{p}; \tvec{t}^{(\sf test)}, \vec{s}_{1:K}^{(\sf test)})$. Note, the clean reference images used in training are not required during inference. We term the training \eqref{our_2} and the corresponding registration procedure as \aimed. 

\begin{figure*}[ht]
    \centering
    \includegraphics[width=0.95\textwidth]{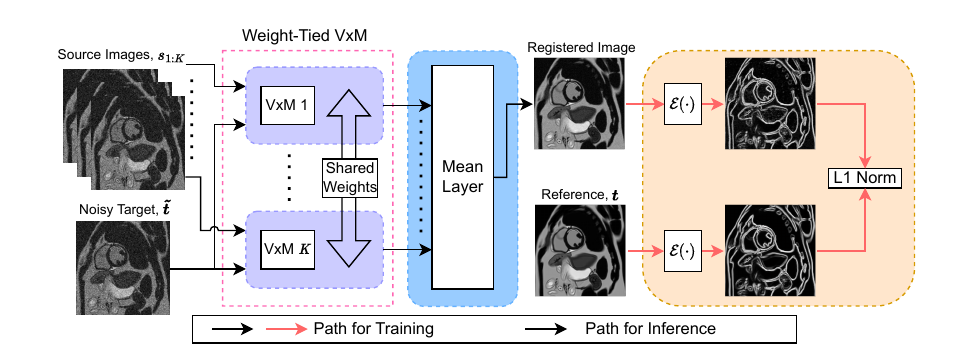}
    \caption{Overview of the proposed \aimed method. One noisy target $\tvec{t}$ and $K$ noisy source images $\vec{s}_{1:K}$ are fed into the network, with each \vxm unit (purple block) receiving $\tvec{t}$ and one of the noisy source images. Each \vxm unit internally generates a deformation field (not shown) and registers one of the source images to $\tvec{t}$. The mean layer averages all the individual registered source images to generate the final registered image. The loss is computed from the clean reference $\vec{t}$ and the final registered image after they both have been passed through a pre-trained edge detector, $\mathcal{E}(\cdot)$. Note, the clean reference image $\vec{t}$ is not needed at the time of inference (black arrows).}
    \label{fig:layout}
\end{figure*}

The training and inference steps in \aimed are depicted in \figref{layout}. A key feature of the architecture employed in \aimed is that it uses $K$ \vxm networks in parallel, followed by a mean layer with no learnable parameters. Since the \vxm networks are weight-tied, \aimed has the same number of learnable parameters (approximately 103,000) as a single \vxm, despite its ability to jointly register multiple source images to a single target image. Being lightweight, \aimed can be trained on smaller training datasets, as demonstrated in the following experimental studies. Another advantage of the architecture used in \aimed is that it can register any number of source images during inference, regardless of the value of $K$ used during training.

\subsection{Study I--LGE phantom}
In this study, realistic free-breathing 2D LGE data were simulated based on the work by Kruithof et al.,\cite{kruithof2021influence} which extends the initial implementation of MRXCAT \cite{wissmann2014mrxcat} to LGE imaging. Single-coil real-valued LGE images from 12 distinct digital patients (DP) were simulated, with each DP characterized by a unique body habitus. Eight short-axis slices were simulated from each DP under free-breathing conditions for a total of 15 ($K+1$) single-shot LGE images per slice. The slice thickness and in-plane resolutions were set at 8 mm and $2 \times 2~\text{mm}^2$, respectively. To generate diversity in respiration-induced image misalignment while maintaining realistic respiratory motion, we modeled the depth, hysteresis, and frequency of respiration based on the work by Lapshin \cite{lapshin1995analytical} and assigned a distinct motion pattern to each DP. Additionally, fibrotic lesions were simulated by altering the relaxation times of a small region in the myocardium. The location and size of the lesion varied across DPs.

\subsection{Study II--LGE from healthy subjects}
Single-shot free-breathing T1-weighted images were collected from 12 healthy volunteers on a 3T scanner (MAGNETOM Vida, Siemens Healthineers, Erlangen, Germany) without contrast. The multi-coil data were collected using phase-sensitive inversion recovery sequence \cite{kellman2002phase} at a prospective acceleration factor of two. For each volunteer, eight short-axis slices were acquired, with 15 ($K+1$) repetitions per slice. The complex-valued images were reconstructed offline using SENSE reconstruction,\cite{pruessmann1999sense} with the sensitivity maps estimated using ESPIRiT.\cite{uecker2014espirit} Other imaging parameters included: spatial resolution $1.48\times 1.48~\text{mm}^2$, slice thickness 8 mm, temporal footprint 205.2 to 212.8 ms, echo time 2.7 to 2.8 ms, inversion time 350 ms, and flip angle 55 degrees.

\subsection{Study III--LGE from patients}
Single-shot free-breathing LGE data were collected from five clinical patients. The data were collected on a {1.5T scanner (MAGNETOM Sola, Siemens Healthineers, Erlangen, Germany)} after administering gadolinium-based contrast. Again, the multi-coil data were collected using the phase-sensitive inversion recovery sequence. From each patient, one short-axis (mid-ventricular) slice was selected, with 15 ($K+1$) repetitions per slice. The complex-valued images were reconstructed offline using SENSE. Other imaging parameters included: spatial resolution $2.2\times 2.2~\text{mm}^2 - 2.5\times 2.5~\text{mm}^2$, slice thickness 8 mm, temporal footprint 242.5 to 297.6 ms, echo time 2.5 to 2.6 ms, inversion time 340 to 420 ms, and flip angle 40 degrees. Three of the five patients were clinically assessed to have positive LGE.

\subsection{Study IV--LGE from low-field scanner}
Six patients were scanned on a 0.55T low-field scanner (MAGNETOM Free.Max, Siemens Healthineers, Erlangen, Germany). In this study, the data were collected with rate-3 prospective undersampling after administering gadolinium-based contrast. Again, the multi-coil data were collected using the phase-sensitive inversion recovery sequence. For each patient, one short-axis slice was acquired, with 15 ($K+1$) repetitions per slice. The complex-valued images were reconstructed offline using SENSE. Other imaging parameters included: spatial resolution $1.8\times 1.8~\text{mm}^2 - 2.1\times 2.1~\text{mm}^2$, echo time 4.2 to 4.4 ms, slice thickness 10 mm, temporal footprint 176.7 to 184.4 ms, inversion time 220 to 260 ms, and flip angle 50 degrees. One of the six patients was clinically assessed to have positive LGE.


\subsection{Implementation details}
All images in Studies II, III, and IV underwent intensity correction using a recently proposed surface coil correction (SCC) method.\cite{lei2023surface} Without intensity correction, cardiac images tend to be brighter in the regions closer to the receive coils and darker toward the middle of the image, which can make image registration challenging in the heart area. We observed SCC to improve the performance of all methods. In addition, all images were cropped to a fixed size of $192 \times 192$ and normalized before being fed into the network. 

In Studies I and II, the 12 subjects were randomly divided into training (6 subjects), validation (3 subjects), and testing (3 subjects) folds. The validation set was used to optimize hyperparameters, including batch size, learning rate, and the regularization parameter $\lambda$. Separate networks were trained for Study I and Study II. However, the network trained in Study II was then applied to the data in Studies III and IV without further training. 

All networks were trained on the same training data using stochastic gradient descent (SGD) with cosine annealing \cite{loshchilov2016sgdr} for $2,\!500$ epochs. {\vxmcc and \vxmed were trained on pairs of images using \eqref{vxm_2}, and \aimcc and \aimed were trained on sets of 15 images (one target and 14 source) using \eqref{our_2}. In all cases, pixelwise mean squared difference was used for $\mathcal{L}_{\sf s}(\cdot, \cdot)$ and $\ell_2$-norm of spatial gradients was used as $\mathcal{L}_{\sf r}(\cdot)$.} To facilitate robust training, data augmentation was applied by dynamically adding varying levels of noise ($7~\text{dB} \pm 3.5~\text{dB}$ for Study I and $11~\text{dB} \pm 3.5~\text{dB}$ for Study II), performing random image shifts, and randomly perturbing image intensity across SGD iterations. The edge detection $\mathcal{E}(\cdot)$ was performed using a pre-trained lightweight dense CNN (LDC).\cite{soria2022ldc} The training was performed on a system equipped with a single Nvidia GeForce RTX 4090 GPU, taking approximately 11 hours to train one model. At inference, the computation time for an image series with $K=14$ was 24 ms. A PyTorch implementation for \aimed can be downloaded from \href{https://github.com/OSU-MR/aimed}{https://github.com/OSU-MR/aimed}. 

To mimic SNRs of 11 dB, 6 dB, and 1 dB during testing, the images in the training folds in Studies I, II, and III were contaminated with three levels of noise. In Study I, the images were real-valued and contaminated with real-valued white Gaussian noise. In Studies II and III, the images were complex-valued and contaminated with complex-valued white Gaussian noise.  The value of 11 dB was selected to mimic the SNR at 1.5T, as observed from a handful of unrelated clinical studies at 1.5T. Likewise, the value of 6 dB was selected to simulate data from a 0.55T scanner. The extreme value of 1 dB was selected as a challenging case to test the limits of various registration methods. Since the SNR calculation and noise addition was performed for the entire dataset in each study, the SNR of individual images varied slightly from the nominal values of 11 dB, 6 dB, and 1 dB.

\subsection{Evaluation}
In each study, we compare the five methods: (i) a traditional nonrigid image registration (\moco) based on minimizing a symmetric energy functional, \cite{guetter2011efficient} (ii) \vxm with CC-based loss (\vxmcc), (iii) \vxm with LDC-based loss (\vxmed), (iv) weight-tied \vxm networks with CC-based loss (\aimcc), and (v) weight-tied \vxm networks with LDC-based loss (\aimed). These methods are summarized in \tabref{methods}. \vxmcc and \vxmed are applied to all $K$ source-target imaging pairs separately, followed by the averaging of the registered images. In contrast, \aimcc and \aimed perform groupwise registration by employing the mean layer, as shown in \figref{layout}. Also, note that \vxmed and \aimcc are included to highlight the independent effects of the groupwise registration and LDC-based loss used in \aimed.  

\begin{table}[!ht]
{
    \centering
    \fontsize{9pt}{12pt}\selectfont
    \begin{tabular}
    {l|c|c}
         \hline
         \multicolumn{1}{l||}{Method}  & Description & Registration \\ \hline
         \multicolumn{1}{l||}{\moco}   & nonrigid energy minimization & pairwise \\
         \multicolumn{1}{l||}{\vxmcc}   & single \vxm + CC loss & pairwise \\
         \multicolumn{1}{l||}{\vxmed}  & single \vxm + LDC loss & pairwise \\
         \multicolumn{1}{l||}{\aimcc}  & weight-tied \vxm\!s + CC loss & groupwise \\
         \multicolumn{1}{l||}{\aimed}  & weight-tied \vxm\!s + LDC loss & groupwise \\ \hline
    \end{tabular}
    \caption{Five image registration methods used in the comparison.}
    \label{tab:methods}}
\end{table}

Although image registration methods based on edge-guidance have been proposed for noncardiac applications,\cite{sideri2022multi,chai2020mri,cao2021edge} the classical gradient-based edge detection used in these methods make them ill-equipped for low SNR conditions. This point is conveyed in Supporting Information \figsupref{edges}, where gradient-based edge detection deteriorates more quickly with noise compared to LDC-based edge detection. Also, our attempts to apply the method by Sideri-Lampretsa et al.,\cite{sideri2022multi} which was originally proposed for high-SNR brain imaging, did not yield reasonable results at the SNR levels considered in this work. Therefore, such methods are not included in the comparison.

To evaluate performance, four image quality metrics were employed to compare the registered image with the clean reference $\vec{t}$. These metrics included recovery SNR (rSNR),  structural similarity index (SSIM), \cite{wang2004image} and two DL-based perceptual metrics, i.e., learned perceptual image patch similarity (LPIPS) \cite{zhang2018unreasonable} and deep image structure and texture similarity (DISTS). \cite{ding2020image} Here, rSNR is defined as the negative of the normalized mean squared error defined in dB. Both rSNR and SSIM are commonly utilized in assessing image quality. The inclusion of LPIPS and DISTS, however, was inspired by a recent study, \cite{kastryulin2023image} where these perceptual quality metrics were shown to correlate with radiologists' subjective assessment of the MRI quality. Since any of the $K+1$ frames can act as a noisy target $\tvec{t}$, the performance metrics were computed and averaged over not only slices and subjects but also $K+1$ possible combinations of $\vec{s}_{1:K}$ and $\tvec{t}$ for each slice. The comparison was repeated at three different values of SNR, i.e., 11 dB, 6 dB, and 1 dB. 

In Study IV, where the high-quality reference was not available, the registered images were blindly scored by three expert readers in terms of overall image quality. Each image was assigned a score on a five-point Likert scale (1: non-diagnostic, 2: poor, 3:
adequate, 4: good, 5: excellent).

\section{Results}\label{sec:res}
In this section, we present results from the four above-mentioned studies. In the first three studies, we compare \moco, \vxmcc, \vxmed, \aimcc, and \aimed in terms of rSNR, SSIM, LPIPS, and DISTS for three different values of SNR. In the last study, we compare these methods in terms of the expert reader scoring.

\begin{figure*}[ht]
\centering
\includegraphics[width=0.95\textwidth]{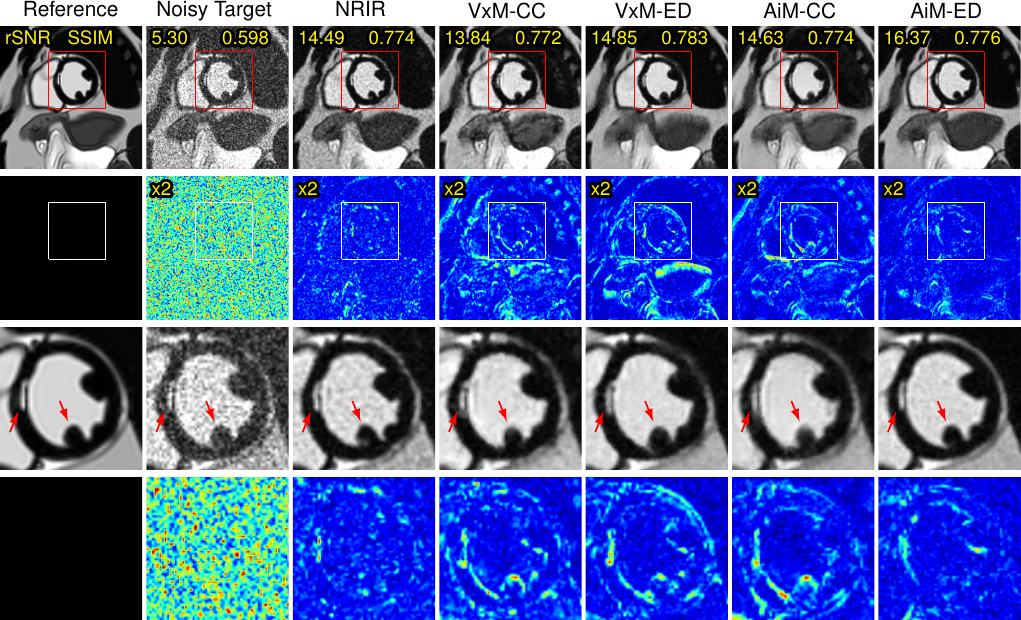}
\includegraphics[width=0.95\textwidth]{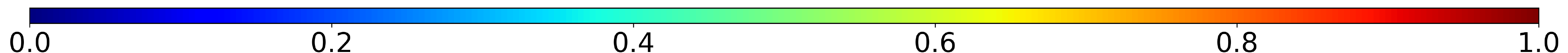}
\caption{Representative results from a digital patient (Study I) with a myocardial lesion at 6 dB SNR. The first row, from left to right, shows the noiseless reference $\vec{t}$, noisy target $\tvec{t}$, and registered images from \moco, \vxmcc, \vxmed, \aimcc, and \aimed. The second row shows the error map with respect to $\vec{t}$ at two-fold amplification. The bottom two rows show a magnified region around the heart, with one red arrow pointing to the lesion and the other pointing to a papillary muscle. }
\label{fig:dp6}
\end{figure*}

\subsection{Study I--LGE phantom}
The four image quality metrics averaged over 360 combinations of target and source images from the test dataset in Study I are summarized in \tabref{rsnr}A. \aimed consistently outperforms other methods in terms of LPIPS and DISTS and is among the top two in terms of rSNR and SSIM. The difference between \aimed and the other methods is more pronounced at lower SNR values. Representative images at 6 dB are shown in \figref{dp6}. Compared to \moco, \aimed is more effective in suppressing noise, and compared to other DL-based methods, \aimed is more effective in preserving small details, as highlighted by the red arrows. Additional representative images at SNRs of 11 dB and 1 dB are shown in \figsupref{dp11} and \figsupref{dp1}, respectively, in the Supporting Information.

\begin{table*}[!ht]
{
\centering
\includegraphics[width=0.99\textwidth]{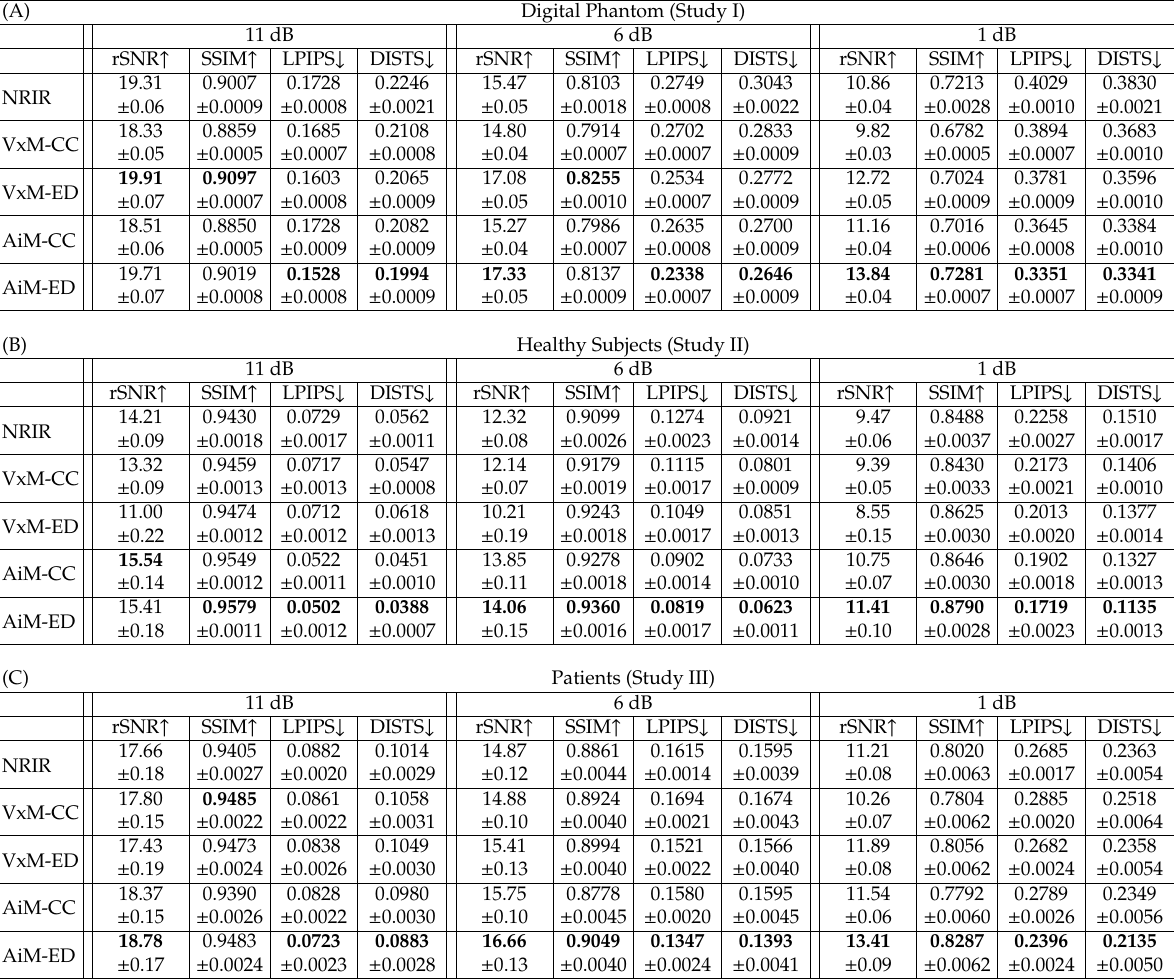}
\caption{Quantitative results (mean$\pm$ standard error) from Studies I, II, and III at three different SNR values. For rSNR and SSIM, higher values are considered better, while for LPIPS and DISTS, lower values are considered better. The best value in each column is highlighted in bold font.}
\label{tab:rsnr}}
\end{table*}

\subsection{Study II--LGE from healthy subjects}
\begin{figure*}[ht]
\centering
\includegraphics[width=0.95\textwidth]{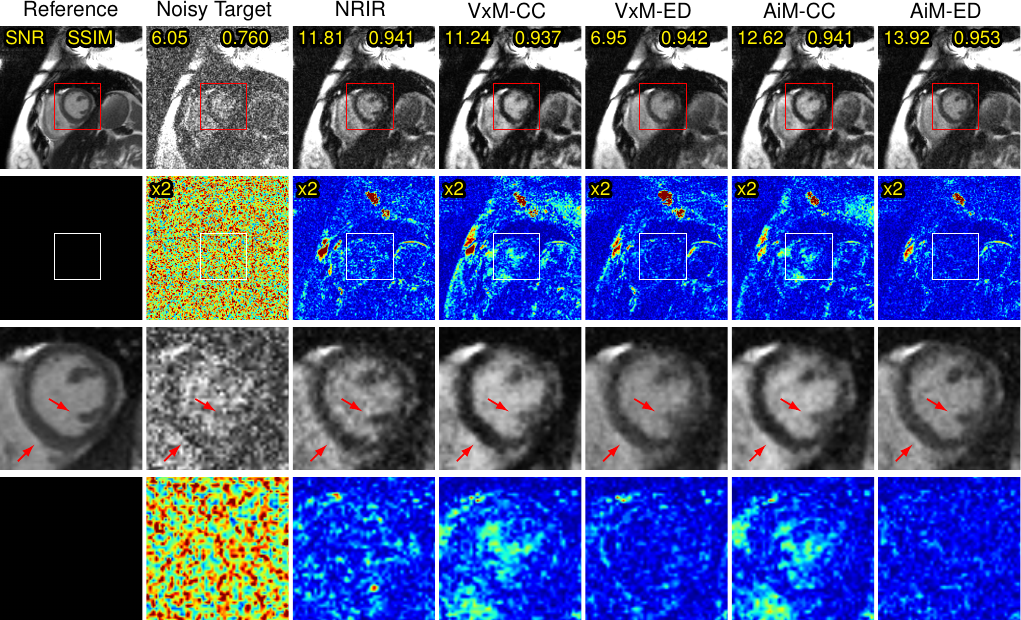}
\includegraphics[width=0.95\textwidth]{colorbar_jet.pdf}
\caption{Representative results from a healthy subject (Study II) at 6 dB SNR. The first row, from left to right, shows the noiseless reference $\vec{t}$, noisy target $\tvec{t}$, and registered images from \moco, \vxmcc, \vxmed, \aimcc, and \aimed. The second row shows the error map with respect to $\vec{t}$ at two-fold amplification. The bottom two rows show a magnified region around the heart, with the red arrow pointing to details lost in some methods. }
\label{fig:hs6}
\end{figure*}

\tabref{rsnr}B provides the values of four image quality metrics averaged over 360 combinations of target and source images from the test dataset in Study II. Again, \aimed consistently outperforms other methods in terms of SSIM, LPIPS, and DISTS and is in the top two in terms of rSNR. In particular, under the low SNR of 1 dB, the rSNR of \aimed is at least 1.94 dB better than the competing methods. A representative image at 6 dB is shown in \figref{hs6}. All image registration methods exhibit degradation in comparison to the noiseless reference. This is, however, not surprising considering the low SNR of the noisy target, particularly in the heart area. Again, compared to \moco, \aimed is more effective in suppressing noise, and compared to other DL-based methods, \aimed is more effective in preserving small details and maintaining the myocardial contrast, as highlighted by the red arrows. In contrast, both \vxmcc and \aimcc have generated images where the myocardium is excessively dark and uneven. Additional representative images at SNRs of 11 dB and 1 dB are shown in \figsupref{hs11} and \figsupref{hs1}, respectively, in the Supporting Information.

\subsection{Study III--LGE from patients}
\begin{figure*}[ht]
\centering
\includegraphics[width=0.95\textwidth]{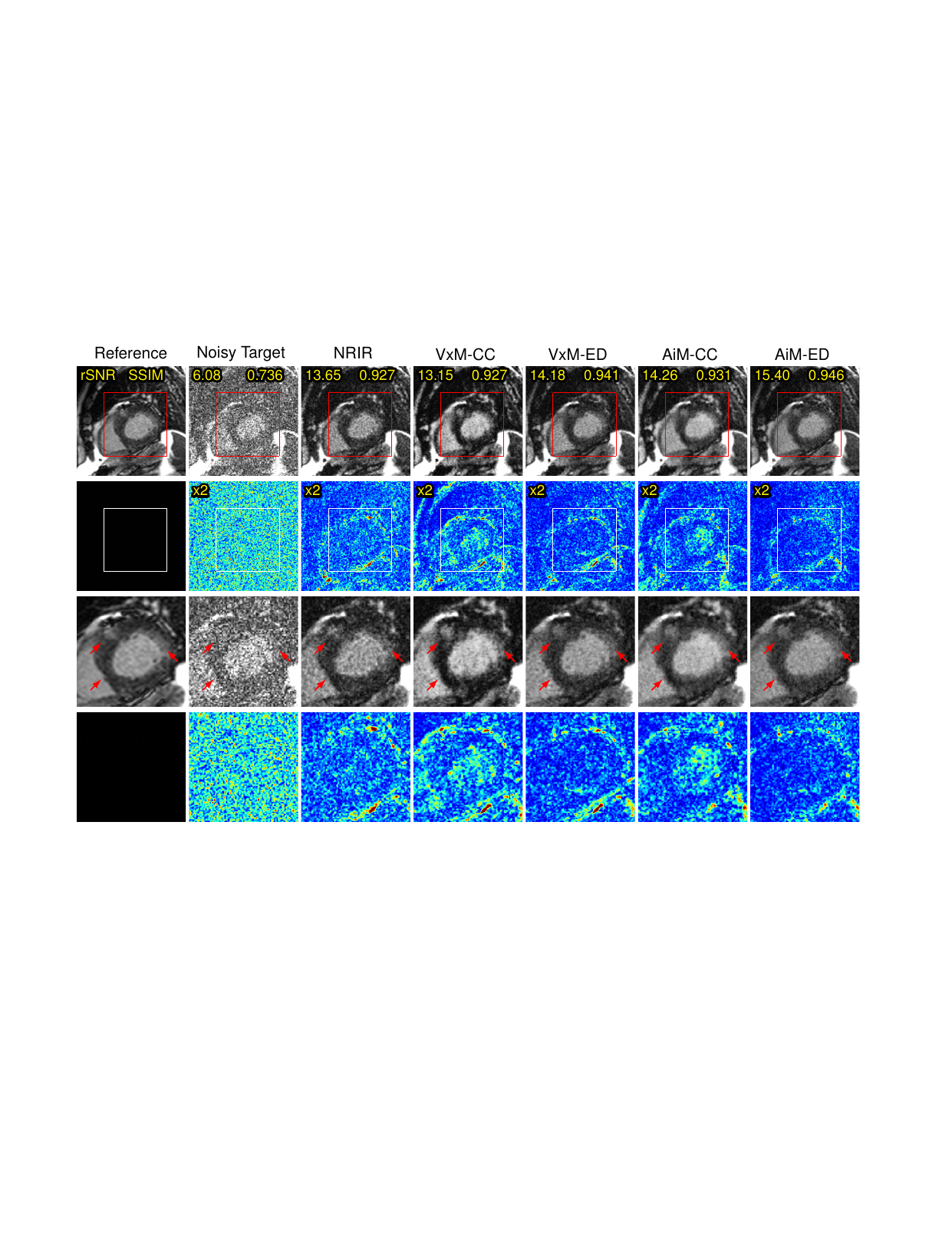}
\includegraphics[width=0.95\textwidth]{colorbar_jet.pdf}
\caption{Representative results from a patient (Study III) at 6 dB SNR. The first row, from left to right, shows the noiseless reference $\vec{t}$, noisy target $\tvec{t}$, and registered images from \moco, \vxmcc, \vxmed, \aimcc, and \aimed. The second row shows the error map with respect to $\vec{t}$ at two-fold amplification. The bottom two rows show a magnified region around the heart. The red arrow on the middle-left points to a focal lesion, the arrow on the middle-right points to diffused enhancement, and the arrow on the bottom-left highlights a healthy myocardium. }
\label{fig:ps6}
\end{figure*}

\tabref{rsnr}C provides the values of four image quality metrics averaged over 75 combinations of target and source images from the test dataset in Study III. The results for this study are consistent with those from Studies I and II, i.e., \aimed consistently outperforms other methods. As observed in Studies I and II, compared to \aimed, some of the competing methods experience a more marked drop in performance from 11 dB to 1 dB. A representative image at 6 dB in \figref{ps6} shows that \aimed is more effective in preserving the shape of the lesion (top two red arrows) and in maintaining the uniformity of the myocardial contrast (bottom red arrow). In comparison, \moco image is excessively grainy, while \vxmcc and, to some extent, \aimcc generate images where the myocardium is excessively dark and uneven. Additional representative images at SNRs of 11 dB and 1 dB are shown in \figsupref{ps11} and \figsupref{ps1}, respectively, in the Supporting Information.

\subsection{Study IV--LGE from low-field scanner}
\begin{figure*}[ht]
\centering
\includegraphics[width=0.95\textwidth]{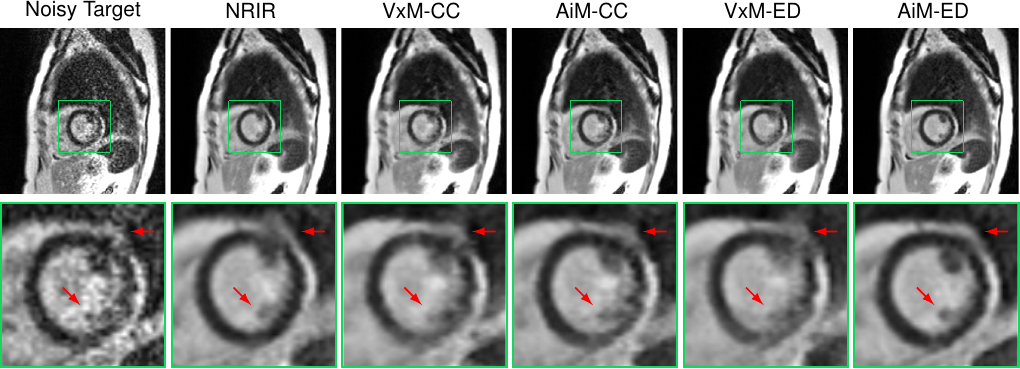}
\caption{Representative results from a patient (Study IV) scanned at 0.55T scanner. The first row, from left to right, shows noisy target $\tvec{t}$, and registered images from \moco, \vxmcc, \vxmed, \aimcc, and \aimed. The second row shows a magnified region around the heart, with the red arrow pointing to areas where loss of detail or artifact is visible.}
\label{fig:0.55T}
\end{figure*}

\tabref{scoring} provides image quality scores from three expert readers for the patient study conducted on the 0.55T scanner. Each number in this table represents an average of over 18 combinations of test and source images. \aimed received the highest score by each of the three readers. \moco and \aimcc have the second and third highest average scores. A representative image is shown in \figref{0.55T}. As was the case in other studies, \aimed preserves more details and exhibits fewer artifacts, as highlighted by the red arrows.

\begin{table}
{
    \centering
    \fontsize{9pt}{11.5pt}\selectfont
    \begin{tabular}
    {l|c|c|c|c}
         \hline
         \multicolumn{1}{l||}{} & ~~~R1~↑~ & ~~~R2~↑~ & ~~~R3~↑~ & ~~Average~↑ \\ \hline
         \multicolumn{1}{l||}{\moco}        & 3.4 $\pm$ 0.3          & 3.6 $\pm$ 0.3           & 3.4 $\pm$ 0.3           & 3.5 $\pm$ 0.2            \\ \hline
         \multicolumn{1}{l||}{\vxmcc}       & 3.0 $\pm$ 0.2          & 2.6 $\pm$ 0.2           & 3.1 $\pm$ 0.2           & 2.9 $\pm$ 0.1           \\ \hline
        \multicolumn{1}{l||}{\vxmed}        & 3.3 $\pm$ 0.3          & 2.5 $\pm$ 0.3           & 2.8 $\pm$ 0.3           & 2.9 $\pm$ 0.2           \\ \hline
         \multicolumn{1}{l||}{\aimcc}       & 3.1 $\pm$ 0.2          & 3.2 $\pm$ 0.2           & 3.5 $\pm$ 0.2           & 3.3 $\pm$ 0.1           \\ \hline
         \multicolumn{1}{l||}{\aimed}       &f{3.5 $\pm$ 0.3}  & f{3.8 $\pm$ 0.3}  & f{3.8 $\pm$ 0.3}  & f{3.7 $\pm$ 0.1} \\ \hline
    \end{tabular}
    \caption{Expert reader scoring of overall image quality in Study IV. Each number represents a mean$\pm$ standard error.}
    \label{tab:scoring}}
\end{table}

\section{Discussion}\label{sec:dis}
We proposed a groupwise image registration method, \aimed, aimed at low-SNR cardiac applications, such as single-shot LGE. Using data from a realistic digital phantom, healthy subjects, and patients across a range of SNR values, we demonstrated the performance advantage of \aimed. In all studies, \aimed consistently outperforms traditional \moco as well as DL-based \vxmcc. Since \aimed is superior to both \aimcc and \vxmed, we conjecture that both the weight-tied \vxm\!s with mean layer and the DL-based edge detection contribute to the superior performance of \aimed. As shown in \figref{dp6} through \figref{ps6}, \aimed provides the best combination of noise suppression and image sharpness while maintaining myocardial homogeneity. The Supporting Information \figsupref{dp11}, \figsupref{hs11}, and \figsupref{ps11} provide representative examples at 11 dB from Studies I, II, and III, respectively. Although all methods yield images of high quality at this SNR, \aimed is more effective in preserving small details. The Supporting Information \figsupref{dp1}, \figsupref{hs1}, and \figsupref{ps1} provide representative examples at 1 dB from Studies I, II, and III, respectively. At this extremely low SNR, the image quality across all methods is marginal and may not be diagnostic. However, \aimed is more effective in preserving myocardial contrast. In the last study, where low-SNR LGE images are collected on a 0.55T scanner, even though the spatial resolution and contrast are different from the 3T images used to train the model, \aimed still outperforms other methods as highlighted by an example in \figref{0.55T}. 

\tabref{rsnr} summarizes the rSNR, SSIM, LPIPS, and DISTS values from Studies I, II, and III for three different values of SNR. As expected, the performances of all methods degrade with lower SNR. However, \aimed is either the best or a close second in all three studies and for all three SNRs. In Study IV, where the reference image was not available, the images were scored blindly by three expert readers. As show in \tabref{scoring}, \aimed receives the highest score from each of the readers.

We attribute the superior performance of \aimed to several factors. First, the use of DL-based edge maps in the loss function enhances robustness against noise and contrast variations. Second, instead of extracting edges from the individual source images, we apply edge detection to the averaged image after the mean layer. Because of averaging, this image has a higher SNR and is thus more amenable to edge detection. Third, in pairwise registration methods, each source image is registered to the noisy target image individually. In the process, each source image partially overfits the noise in the target image. As a result, all registered source images inherit some level of noise from the target image. Since this noise is correlated across all registered source images, the post-registration averaging becomes less effective. This is less of a problem for \vxmed because it focuses mostly on edges and not on noise or texture. In contrast, \aimed and \aimcc do not fit individual source images to the target, leading to better noise suppression. Finally, the lightweight nature of \aimed, achieved by employing weight-tied \vxm networks, allows for efficient training and inference without the need for large datasets. This is particularly advantageous in clinical environments where high-quality reference images are often unavailable, especially for cardiac applications. Another practical advantage of \aimed, compared to \moco or similar methods, is computation speed. At the inference stage, \aimed could register 14 source images to a noisy target image in approximately 24 ms, compared to 3.96 s for \moco.

Despite the promising initial results, \aimed has limitations that require further investigation. First, the current study focuses on 2D image registration, and future work should explore its extension to 3D imaging. Second, we have exclusively focused on single-shot LGE applications, but other applications such as mapping could also benefit from these developments and should be explored. Third, being a 2D method, \aimed cannot correct for the through-plane motion. However, there are possible extensions that can minimize the impact of through-plane motion. For example, one could replace the mean layer with a weighted averaging layer, where the weights are inversely related to the $\ell_2$-norm of the difference between the registered source images from their mean. This way, if a specific registered source image is further away from the mean due to the through-plane motion, it contributes less to the final averaged image. More broadly, one could replace the mean layer with a network that combines different source images in a more complex manner. Finally, we have validated \aimed solely as a post-reconstruction image registration method, but its integration with image reconstruction methods should be further investigated.

\section{Conclusion}\label{sec:con}
In this work, we introduced \aimed, a novel groupwise image registration method. The unique integration of DL-based edge detection and a lightweight architecture enhances its accuracy and robustness, especially in challenging imaging scenarios of low SNR. Comprehensive evaluations using data from a digital phantom, healthy subjects, and patients demonstrated that \aimed consistently outperforms traditional and DL-based methods in terms of rSNR, SSIM, LPIPS, and DISTS metrics. Fast computation, ability to handle a wide range of SNRs, and insensitivity to changes in contrast and resolution make \aimed a promising tool for cardiac applications.

\section*{Acknowledgments}
We thank Yingmin Liu for his assistance with data acquisition and Ning Jin from Siemens Healthineers for providing the \moco code.

\subsection*{Financial disclosure}
The authors have no relevant financial disclosures.

\subsection*{Data Availability Statement}
The Python implementation of the AiM-ED method is available on GitHub at \href{https://github.com/OSU-MR/aimed}{https://github.com/OSU-MR/aimed}.

\subsection*{Conflict of interest}
The authors declare no conflict of interest.

\bibliography{root.bib}%

\clearpage
\newpage
\section*{Supporting Information}\label{sec:supp}
The following supporting information is available as part of the online article:

\noindent 
\figsupref{edges}: Edge detection applied across six different levels of SNR. Top row: a single frame from the digital phantom described in Study I contaminated with various levels of white Gaussian noise. Middle row: corresponding edges from the DL-based LDC edge detector. Bottom row: corresponding edges from a classical gradient-based edge detection used in the work by Sideri-Lampretsa et al.\cite{sideri2022multi}

\noindent
\figsupref{dp11}: Representative results from a digital patient (Study I) with a myocardial lesion at 11 dB SNR. The first row, from left to right, shows the noiseless reference $\vec{t}$, noisy target $\tvec{t}$, and registered images from \moco, \vxmcc, \vxmed, \aimcc, and \aimed. The second row shows the error map with respect to $\vec{t}$ at two-fold amplification. The bottom two rows show a magnified region around the heart, with one red arrow pointing to the lesion and the other pointing to the papillary muscle.

\noindent
\figsupref{dp1}: Representative results from a digital patient (Study I) with a myocardial lesion at 1 dB SNR. The first row, from left to right, shows the noiseless reference $\vec{t}$, noisy target $\tvec{t}$, and registered images from \moco, \vxmcc, \vxmed, \aimcc, and \aimed. The second row shows the error map with respect to $\vec{t}$ at two-fold amplification. The bottom two rows show a magnified region around the heart, with one red arrow pointing to the lesion and the other pointing to the papillary muscle.

\noindent
\figsupref{hs11}: Representative results from a healthy subject (Study II) at 11 dB SNR. The first row, from left to right, shows the noiseless reference $\vec{t}$, noisy target $\tvec{t}$, and registered images from \moco, \vxmcc, \vxmed, \aimcc, and \aimed. The second row shows the error map with respect to $\vec{t}$ at two-fold amplification. The bottom two rows show a magnified region around the heart, with the red arrow pointing to details lost in some methods.

\noindent
\figsupref{hs1}: Representative results from a healthy subject (Study II) at 1 dB SNR. The first row, from left to right, shows the noiseless reference $\vec{t}$, noisy target $\tvec{t}$, and registered images from \moco, \vxmcc, \vxmed, \aimcc, and \aimed. The second row shows the error map with respect to $\vec{t}$ at two-fold amplification. The bottom two rows show a magnified region around the heart, with the red arrow pointing to details lost in some methods. 

\noindent
\figsupref{ps11}: Representative results from a patient (Study III) at 11 dB SNR. The first row, from left to right, shows the noiseless reference $\vec{t}$, noisy target $\tvec{t}$, and registered images from \moco, \vxmcc, \vxmed, \aimcc, and \aimed. The second row shows the error map with respect to $\vec{t}$ at two-fold amplification. The bottom two rows show a magnified region around the heart. The red arrow on the middle-left points to a focal lesion, the arrow on the middle-right points to diffused enhancement, and the arrow on the bottom-left highlights a healthy myocardium. 

\noindent
\figsupref{ps1}: Representative results from a patient (Study III) at 1 dB SNR. The first row, from left to right, shows the noiseless reference $\vec{t}$, noisy target $\tvec{t}$, and registered images from \moco, \vxmcc, \vxmed, \aimcc, and \aimed. The second row shows the error map with respect to $\vec{t}$ at two-fold amplification. The bottom two rows show a magnified region around the heart. The red arrow on the middle-left points to a focal lesion, the arrow on the middle-right points to diffused enhancement, and the arrow on the bottom-left highlights a healthy myocardium.

\begin{suppfigure*}[p]
    \flushleft
    \begin{picture}(252,252)
        \put(8,162){\includegraphics[width=0.965\textwidth]{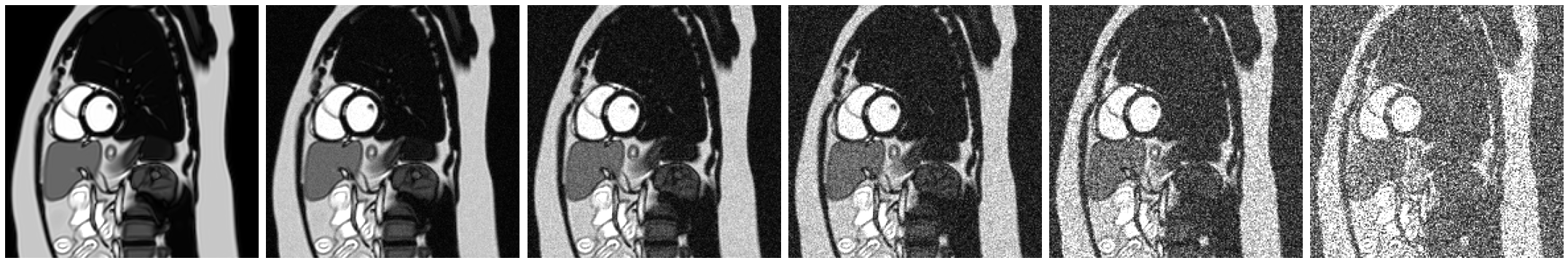}}
        \put(8,81){\includegraphics[width=0.965\textwidth]{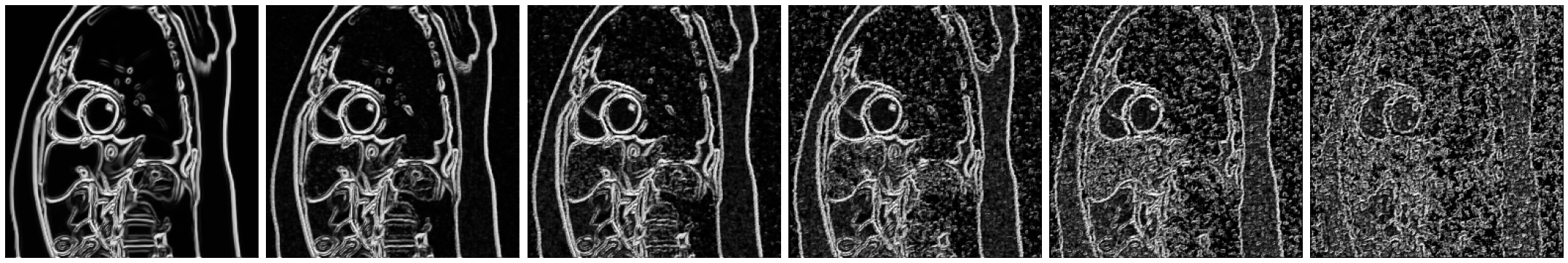}}
        \put(8,0){\includegraphics[width=0.965\textwidth]{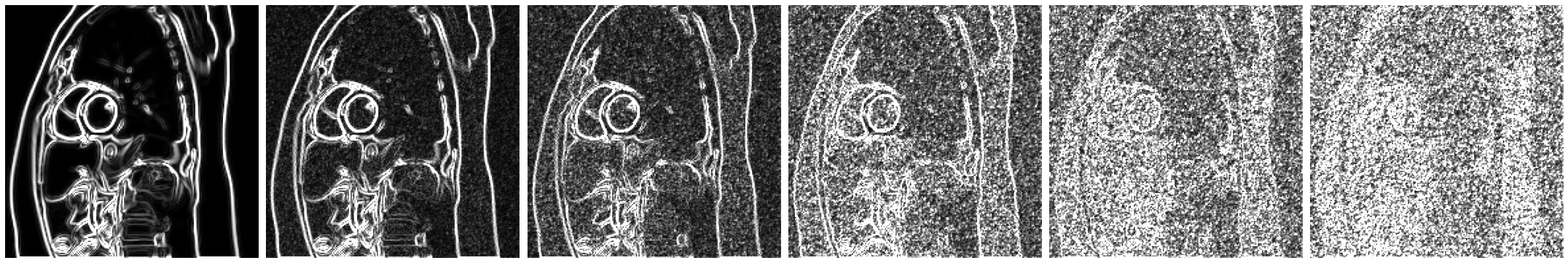}}
        \put(9,236.8){\makebox(0,0)[l]{\sffamily \contour{black}{\textcolor{yellow}{\imagetextsize \hspace*{0em} No Noise}}}}
        \put(91,236.8){\makebox(0,0)[l]{\sffamily \contour{black}{\textcolor{yellow}{\imagetextsize \hspace*{0em} 23 dB}}}}
        \put(173,236.8){\makebox(0,0)[l]{\sffamily \contour{black}{\textcolor{yellow}{\imagetextsize \hspace*{0em} 16 dB}}}}
        \put(255,236.8){\makebox(0,0)[l]{\sffamily \contour{black}{\textcolor{yellow}{\imagetextsize \hspace*{0em} 11 dB}}}}
        \put(337,236.8){\makebox(0,0)[l]{\sffamily \contour{black}{\textcolor{yellow}{\imagetextsize \hspace*{0em} 6 dB}}}}
        \put(419,236.8){\makebox(0,0)[l]{\sffamily \contour{black}{\textcolor{yellow}{\imagetextsize \hspace*{0em} 1 dB}}}}
    \end{picture}
    \caption{Edge detection applied across five different levels of SNR. Top row: a single frame from the digital phantom described in Study I contaminated with various levels of white Gaussian noise. Middle row: corresponding edges from the DL-based LDC edge detector. Bottom row: corresponding edges from a classical gradient-based edge detection used in work by Sideri-Lampretsa et al.\cite{sideri2022multi}}
    \label{fig:edges}
\end{suppfigure*}
\newpage

\begin{suppfigure*}[ht]
\centering
\includegraphics[width=0.95\textwidth]{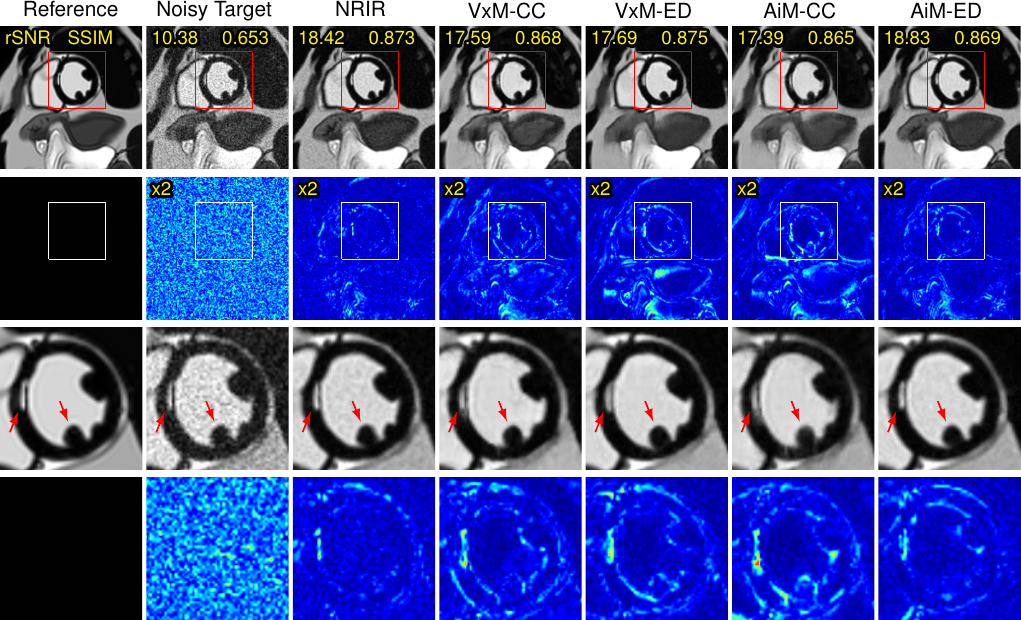}
\includegraphics[width=0.95\textwidth]{colorbar_jet.pdf}
\caption{Representative results from a digital patient (Study I) with a myocardial lesion at 11 dB SNR. The first row, from left to right, shows the noiseless reference $\vec{t}$, noisy target $\tvec{t}$, and registered images from \moco, \vxmcc, \vxmed, \aimcc, and \aimed. The second row shows the error map with respect to $\vec{t}$ at two-fold amplification. The bottom two rows show a magnified region around the heart, with one red arrow pointing to the lesion and the other pointing to the papillary muscle. }
\label{fig:dp11}
\end{suppfigure*}
\newpage

\begin{suppfigure*}[ht]
\centering
\includegraphics[width=0.95\textwidth]{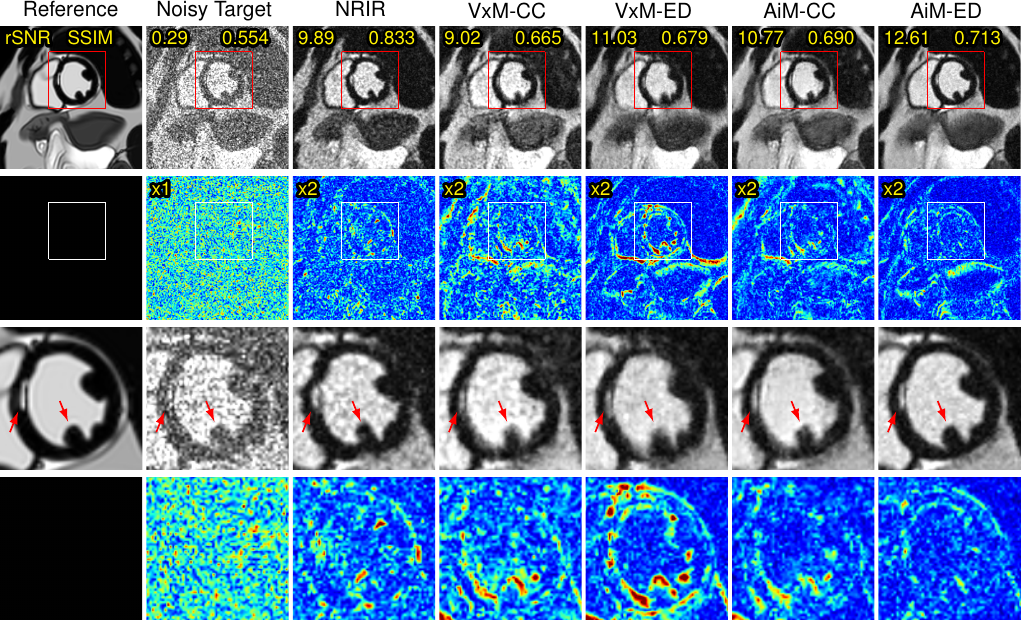}
\includegraphics[width=0.95\textwidth]{colorbar_jet.pdf}
\caption{Representative results from a digital patient (Study I) with a myocardial lesion at 1 dB SNR. The first row, from left to right, shows the noiseless reference $\vec{t}$, noisy target $\tvec{t}$, and registered images from \moco, \vxmcc, \vxmed, \aimcc, and \aimed. The second row shows the error map with respect to $\vec{t}$ at two-fold amplification. The bottom two rows show a magnified region around the heart, with one red arrow pointing to the lesion and the other pointing to the papillary muscle. }
\label{fig:dp1}
\end{suppfigure*}
\newpage

\begin{suppfigure*}[ht]
\centering
\includegraphics[width=0.95\textwidth]{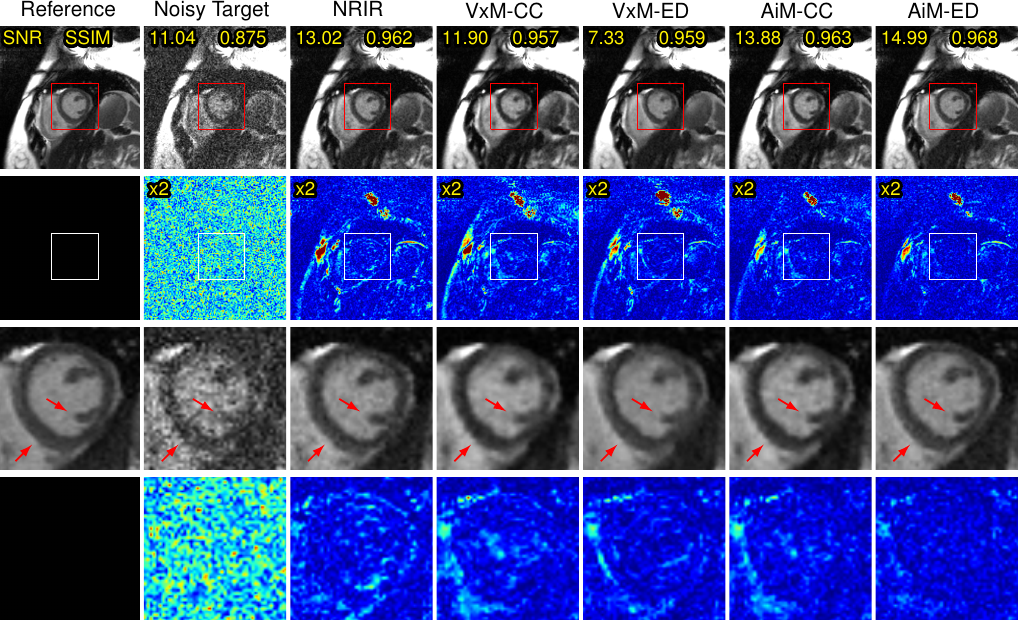}
\includegraphics[width=0.95\textwidth]{colorbar_jet.pdf}
\caption{Representative results from a healthy subject (Study II) at 11 dB SNR. The first row, from left to right, shows the noiseless reference $\vec{t}$, noisy target $\tvec{t}$, and registered images from \moco, \vxmcc, \vxmed, \aimcc, and \aimed. The second row shows the error map with respect to $\vec{t}$ at two-fold amplification. The bottom two rows show a magnified region around the heart, with the red arrow pointing to details lost in some methods. }
\label{fig:hs11}
\end{suppfigure*}
\newpage

\begin{suppfigure*}[ht]
\centering
\includegraphics[width=0.95\textwidth]{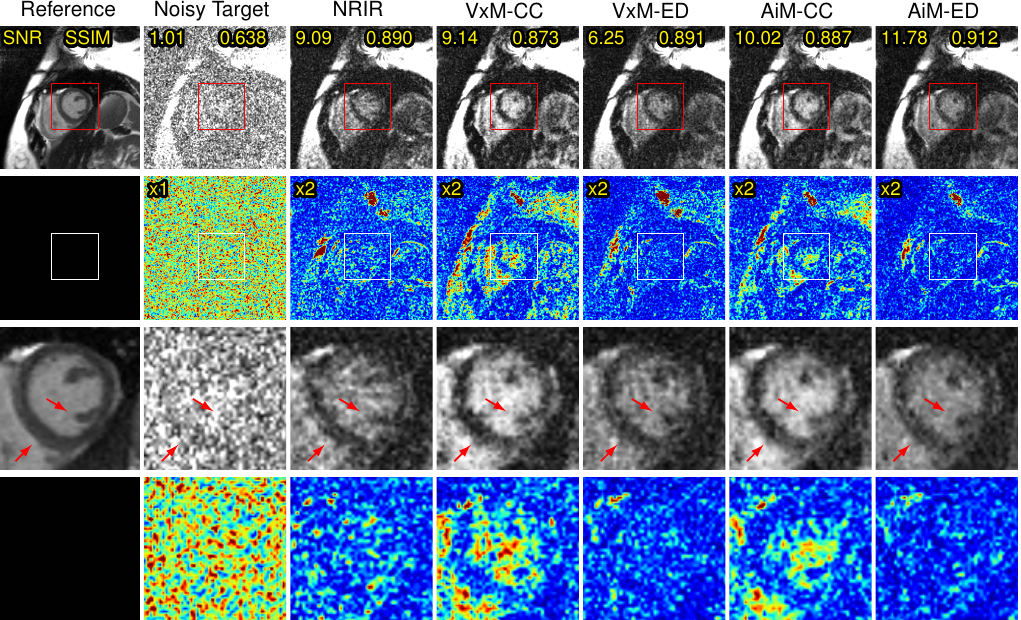}
\includegraphics[width=0.95\textwidth]{colorbar_jet.pdf}
\caption{Representative results from a healthy subject (Study II) at 1 dB SNR. The first row, from left to right, shows the noiseless reference $\vec{t}$, noisy target $\tvec{t}$, and registered images from \moco, \vxmcc, \vxmed, \aimcc, and \aimed. The second row shows the error map with respect to $\vec{t}$ at two-fold amplification. The bottom two rows show a magnified region around the heart, with the red arrow pointing to details lost in some methods. }
\label{fig:hs1}
\end{suppfigure*}
\newpage

\begin{suppfigure*}[ht]
\centering
\includegraphics[width=0.95\textwidth]{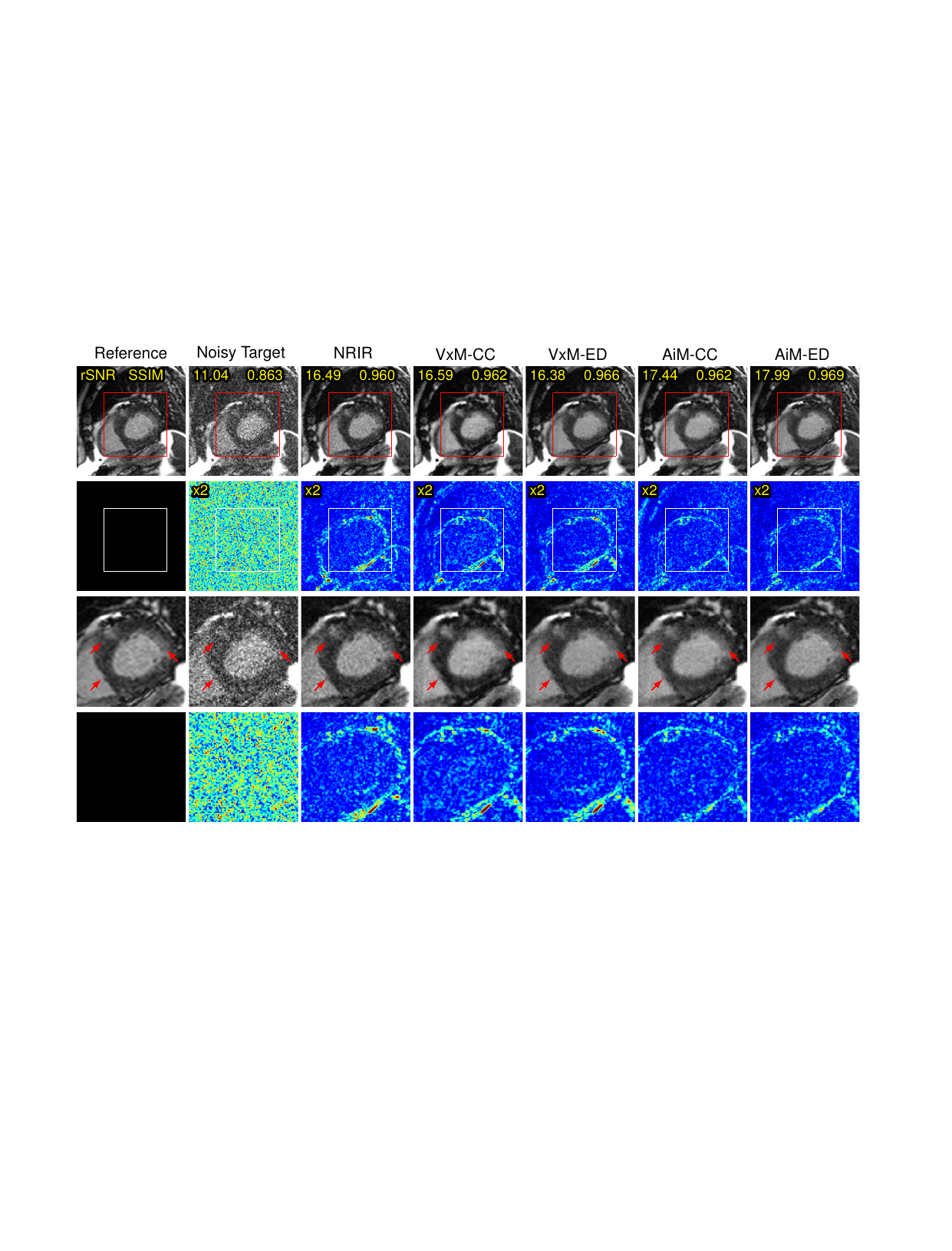}
\includegraphics[width=0.95\textwidth]{colorbar_jet.pdf}
\caption{Representative results from a patient (Study III) at 11 dB SNR. The first row, from left to right, shows the noiseless reference $\vec{t}$, noisy target $\tvec{t}$, and registered images from \moco, \vxmcc, \vxmed, \aimcc, and \aimed. The second row shows the error map with respect to $\vec{t}$ at two-fold amplification. The bottom two rows show a magnified region around the heart. The red arrow on the middle-left points to a focal lesion, the arrow on the middle-right points to diffused enhancement, and the arrow on the bottom-left highlights a healthy myocardium. }
\label{fig:ps11}
\end{suppfigure*}
\newpage

\begin{suppfigure*}[ht]
\centering
\includegraphics[width=0.95\textwidth]{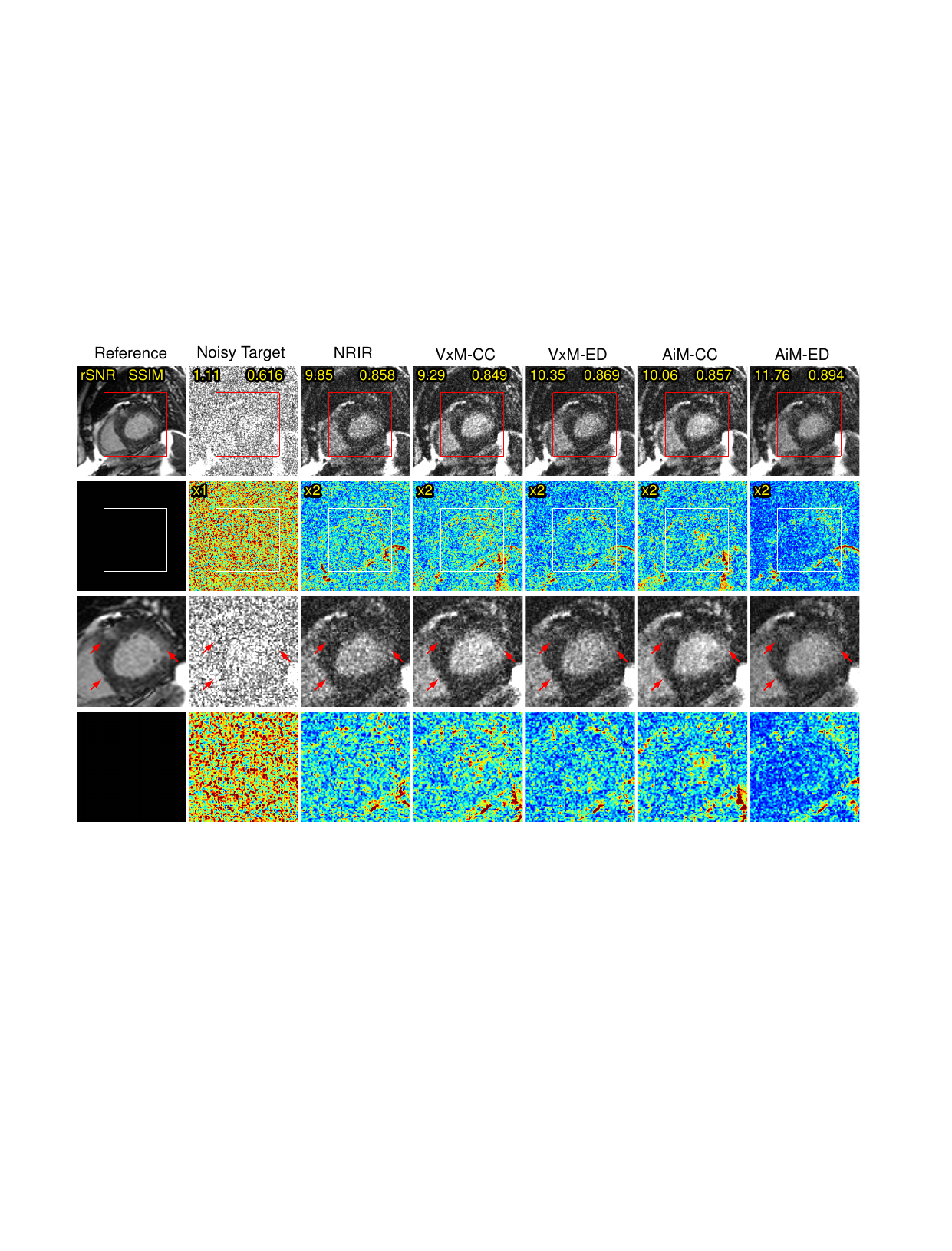}
\includegraphics[width=0.95\textwidth]{colorbar_jet.pdf}
\caption{Representative results from a patient (Study III) at 1 dB SNR. The first row, from left to right, shows the noiseless reference $\vec{t}$, noisy target $\tvec{t}$, and registered images from \moco, \vxmcc, \vxmed, \aimcc, and \aimed. The second row shows the error map with respect to $\vec{t}$ at two-fold amplification. The bottom two rows show a magnified region around the heart. The red arrow on the middle-left points to a focal lesion, the arrow on the middle-right points to diffused enhancement, and the arrow on the bottom-left highlights a healthy myocardium. }
\label{fig:ps1}
\end{suppfigure*}

\end{document}